\documentclass[preprintnumbers, floatfix, preprintnumbers, superscriptaddress, letterpaper, nofootinbib, twocolumn]{revtex4}
\pdfoutput=1
\usepackage{graphicx}
\usepackage{microtype}
\usepackage{amsmath}
\usepackage{amssymb}
\usepackage{subfigure}
\usepackage{hyperref}
\usepackage{url}
\usepackage{xcolor}
\usepackage{color}
\usepackage{mathrsfs}
\usepackage{calrsfs}
\usepackage{amsfonts}
\usepackage{latexsym}
\usepackage{ragged2e}
\usepackage{epsfig}
\usepackage{textcomp}
\usepackage{phaistos}
\usepackage{physics}

\makeatletter
\renewcommand\@makefnmark{\hbox{\@textsuperscript{\normalfont\color{purple}\@thefnmark}}}
\renewcommand\@makefntext[1]{%
\parindent 1em\noindent
\hb@xt@1.8em{%
\hss\@textsuperscript{\normalfont\@thefnmark}}#1}
\makeatother

\usepackage{caption}
\DeclareCaptionJustification{justified}{\leftskip=0pt \rightskip=0pt \parfillskip=0pt plus 1fil}
\definecolor{vividviolet}{rgb}{0.62, 0.0, 1.0}
\definecolor{amaranth}{rgb}{0.9, 0.17, 0.31}
\definecolor{palatinateblue}{rgb}{0.15, 0.23, 0.89}
\definecolor{brightpink}{rgb}{1.0, 0.0, 0.5}
\definecolor{cornflowerblue}{rgb}{0.39, 0.58, 0.93}
\definecolor{deepcarminepink}{rgb}{0.94, 0.19, 0.22}
\definecolor{radicalred}{rgb}{1.0, 0.21, 0.37}

\hypersetup{ linktoc=all,
colorlinks, linkcolor={palatinateblue},
citecolor={brightpink}, urlcolor={black}
}

\graphicspath{{Images/}}

\renewcommand{\d}[1]{\ensuremath{\operatorname{d}\!{#1}}}

\graphicspath{{Images/}}
\renewcommand{\d}[1]{\ensuremath{\operatorname{d}\!{#1}}}


\def\sideremark#1{\ifvmode\leavevmode\fi\vadjust{\vbox to0pt{\vss
\hbox to 0pt{\hskip\hsize\hskip1em
\vbox{\hsize1.5cm\tiny\raggedright\pretolerance10000
\noindent #1\hfill}\hss}\vbox to8pt{\vfil}\vss}}}%
\begin{document}

\title{Measurement-Induced Perturbations of Hausdorff Dimension in Quantum Paths}

\author{You-Wei Ding}
\email{dingyw512@outlook.com}
\affiliation{Center for Gravitation and Cosmology, College of Physical Science and Technology, Yangzhou University, \\180 Siwangting Road, Yangzhou City, Jiangsu Province 225002, China}

\author{Yen Chin \surname{Ong}}
\email{ongyenchin@nuaa.edu.cn}
\affiliation{Center for the Cross-disciplinary Research of Space Science and Quantum-technologies (CROSS-Q),\\
College of Physics, Nanjing University of Aeronautics and Astronautics,\\
29 Jiangjun Road, Nanjing City, Jiangsu Province 211106, China}
\affiliation{Center for Gravitation and Cosmology, College of Physical Science and Technology, Yangzhou University, \\180 Siwangting Road, Yangzhou City, Jiangsu Province 225002, China}

\author{Hao Xu}
\thanks{Corresponding author}
\email{haoxu@yzu.edu.cn}
\affiliation{Center for Gravitation and Cosmology, College of Physical Science and Technology, Yangzhou University, \\180 Siwangting Road, Yangzhou City, Jiangsu Province 225002, China}

\begin{abstract}
{
In a seminal paper, Abbott et al. analyzed the relationship between a particle’s trajectory and the resolution of position measurements performed by an observer at fixed time intervals. They predicted that quantum paths exhibit a universal Hausdorff dimension that transitions from $d=2$ to $d=1$ as the momentum of the particle increases. However, although measurements were assumed to occur at intervals of time, the calculations only involved evaluating the expectation value of operators for the free evolution of wave function within a single interval, with no actual physical measurements performed.} In this work we investigate how quantum measurements alter the fractal geometry of quantum particle paths. By modelling sequential measurements using Gaussian wave packets for both the particle and the apparatus, we reveal that the dynamics of the measurement change the roughness of the path and shift the emergent Hausdorff dimension towards a lower value in nonselective evolution. For selective evolution, feedback control forces must be introduced to counteract stochastic wave function collapse, stabilising trajectories and enabling dimensionality to be tuned. {When the contribution of the measurement approaches zero, our result reduces to that of Abbott et al. Our work can thus be regarded as a more realistic formulation of their approach, and it connects theoretical quantum fractality with measurement physics, quantifying how detectors reshape spacetime statistics at quantum scales.}
\end{abstract}

\maketitle

\section{Introduction}

Self-similarity---a structural property where patterns repeat identically across scales---reveals fundamental physical principles. This scale invariance, characterized by symmetry under scale transformations, manifests across diverse systems including fractal geometries, turbulent flows, and critical phenomena near phase transitions \cite{hutchinson1981fractals,Benzi:1993,Yukalov:2019dap}. It provides a framework for identifying emergent order in complex systems through recursive modeling, notably underpinning renormalization group theory which explains universality.

Scale invariance also plays a crucial role in gravity and high-energy physics, often extending to broader symmetries like conformal invariance. Conformal gravity and conformal field theory (CFT) embody conformal symmetry by remaining invariant under conformal transformations \cite{Mannheim:2011ds,francesco2012conformal}. Furthermore, conformal symmetry is pivotal in the study of quantum gravity. According to the AdS/CFT correspondence, the gravitational theory in anti-de Sitter (AdS) spacetime is mathematically equivalent to a scale-invariant CFT on the low-dimensional boundary of that spacetime \cite{Maldacena:1997re,Witten:1998qj}. This duality establishes a connection between scale-invariant quantum field theories and gravitational systems.

Given the pervasiveness of self-similarity, particularly at the quantum gravity frontier as exemplified by AdS/CFT, it is possible that spacetime itself possesses intrinsic self-similar structures. 
This geometric scale invariance could become evident in quantum particle trajectories and statistical behaviour, which could in turn give rise to critical dynamics or emergent fractal signatures. It should be noted that in some quantum gravity theories, the effective dimensions of spacetime are dynamical and energy scale-dependent \cite{Carlip:2017eud}. The effective dimensions could be either the spectral dimension, or the ``thermal dimension'' \cite{1602.08020,2009.08556}. The possibility of black hole horizons being fractal-like has also been proposed in the context of Barrow entropy \cite{2004.09444}, which could also be energy scale-dependent \cite{2205.09311}. Prior to Barrow entropy, the possibility that the horizon of a black hole becomes strongly ``wrinkled'' and manifesting fractal-like structures was already proposed by Sorkin in \cite{9701056}. 

Despite the aforementioned motivations, it should be emphasized that even before quantum gravity effects are considered, fractal geometry can already arise in the context of quantum particle trajectories. This should be fully understood before we could hope to extract any information unique to quantum gravity. Let us start by summarizing the results of \cite{Abbott:1979bh}, in which the authors establish that one-dimensional quantum-mechanical particle paths exhibit fractal geometry with a Hausdorff dimension of 2. The picture is as follows: one is supposed to ``measure'' the position of a free particle with a fixed spatial resolution $\Delta x$ at times separated by an interval $t$. Thus, starting from some $t_0$, and $t_1=t_0+t, \cdots, t_N=t_0+Nt$, we have the total time $T=t_N-t_0=Nt$. The average length measured would be $\langle l \rangle = N \langle \Delta l \rangle $.
Similar to mathematical constructs such as the Koch snowflake, the measured path length diverges with improved spatial resolution according to the equation:
\( \langle l \rangle \propto {\hbar t}/{m \Delta x} \propto {\hbar T}/{m \Delta x} \)
due to Heisenberg's uncertainty principle\footnote{Essentially, one expects that $\langle l \rangle \propto \Delta x \propto \hbar/\Delta p=\hbar/m\Delta v$. For a more rigorous treatment, see the next section.}. The resolution-independent Hausdorff length 
\begin{align}
\langle L \rangle := \langle l \rangle (\Delta x)^{d-1}
\label{eq1}
\end{align}
yields \( d = 2 \) when \( \langle L \rangle \) remains constant\footnote{As discussed in \cite{Abbott:1979bh}, this is exactly the same as usual fractal geometries such as the Koch snowflake. To see that, suppose that we measure the length of the Koch snowflake to be $l$ at resolution $\Delta x$. If we then increase the resolution to the finer $\Delta x'=(1/3)\Delta x$, then we can measure the next iteration $l'=(4/3)l$ of the snowflake (whose side length increases by a factor of $4/3$ per iteration). The Hausdorff length $L=l(\Delta x)^{d-1}$, is required to be independent of $\Delta x$, so $L'=L$, which implies $d=\ln 4/\ln 3 \approx 1.26186$. The only difference is that in quantum mechanics the path length is an averaged quantity, hence denoted by \( \langle L \rangle \).} as \( \Delta x \to 0 \). Here $l$ is the usual length measured when the resolution is simply $\Delta x$. 
Self-similarity (i.e. $\langle \Delta L \rangle \propto \Delta x$) emerges under the scaling condition $t \propto m (\Delta x)^2 / \hbar$, which reflects the kinetic energy relation \( E = p^2 / 2m \) via the energy-time uncertainty relation. For particles with average momentum \( p_{\text{av}} \), the dimension transitions from classical \( d = 1 \) (when \( \Delta x \gg \hbar / |p_{\text{av}}| \)) to quantum \( d = 2 \) (when \( \Delta x \ll \hbar / |p_{\text{av}}| \)).

It is perhaps worth mentioning that the fractal dimension of Brownian motion is also known to be equal to 2 in dimensions $d \geqslant 2$ \cite{Kalia,1107.3922}. Given the discussions above, this result may not be so surprising since the Brownian motion is described by the heat equation (more generally the Fokker–Planck equation), which is related to the Schr\"odinger equation via a Wick rotation \cite{Favella}. 

{
Returning to the fundamental question of the fractal dimension of quantum mechanical paths, we note that while the original work \cite{Abbott:1979bh} provides valuable insight by analyzing the expected path length via periodic position measurements, it conceptualizes measurements in an abstract manner—reducing them to mathematical operations that yield expectation values such as \( \langle |x(t)| \rangle \), without performing concrete measurement calculations or addressing how the quantum state of the system evolves post-measurement. Although mathematically convenient, this approach represents a significant oversimplification from an experimental perspective.

In real experimental scenarios, measurements are not passive calculations; they involve physical interactions between the quantum system and a measuring apparatus. These interactions inevitably introduce decoherence and wave function collapse, both of which dynamically alter the system's state. The idealized model neglects the essential physics of measurement backreaction, which perturbs the trajectories of quantum particles and modifies the statistical properties of the path ensemble. Thus, while the model correctly captures intrinsic quantum-statistical divergences (\( \langle l \rangle \propto 1/\Delta x \)), it fails to account for measurement-induced disturbances that are unavoidable in practice.}

Moving beyond idealized measurement paradigms, this work explicitly models quantum measurement dynamics by representing both the particle and apparatus as Gaussian wavepackets. We incorporate a dynamically coupled interaction Hamiltonian to mediate (dis)continuous measurement processes, where instrumental backreaction could be used to induce controlled wave function collapse \cite{Caves:1987}. This integrated approach systematically quantifies how detector-induced decoherence, feedback control dynamics, and measurement-conditioned evolution collectively reshape path roughness and alter the emergent Hausdorff dimensionality beyond the nominal \( d = 2 \) value. This in turn paves the road for future experimental quantum measurement physics that can further investigate the notion of fractal dimensions in quantum mechanics.

The remainder of the paper is organized as follows. In Section 2, we provide a concise review of self-similarity and Hausdorff dimensionality in quantum particle paths. In Section 3, we construct the interaction between particles and measuring devices, analyse non-selective measurements in the continuous limit, and introduce feedback control for selective measurements to correct particle paths and dimensions. In Section 4 we summarize the results and suggest some possible directions for future research.

\section{Fractal Dimension of a quantum-mechanical path}

In this section, we review key concepts regarding quantum free particle trajectories developed in \cite{Abbott:1979bh}. Consider a particle evolving over total time \( T \), with positions recorded at intervals \( t \). The average displacement per each measurement is \( \langle \Delta l \rangle \), giving total (averaged) path length
\begin{equation}
\langle l \rangle = \frac{T}{t} \langle \Delta l \rangle,
\end{equation}
from which we derive the Hausdorff dimension using \eqref{eq1}.

This fundamental relationship connects position resolution \( \Delta x \) and path length \( \langle l \rangle \). Any position-space wave function admits a Fourier representation
\begin{equation}
\psi(x) = \frac{1}{\sqrt{2\pi\hbar}} \int_{-\infty}^{\infty} \d p ~\phi(p) e^{ipx/\hbar} ,
\end{equation}
where \( \phi(p) \) denotes the momentum-space wave function. Introducing the dimensionless coefficient \( k \equiv p \Delta x / \hbar \), we require separable dependence on \( k \) and \( \Delta x \) so that \( \phi(p) = f(k) (\Delta x / \hbar)^{1/2} \). This also means that we impose restrictions on the choice of wave functions and $\Delta x$. Consider a Gaussian wave packet as an example:
\begin{equation}
\phi(p) = \left( \frac{\Delta}{\pi \hbar^2} \right)^{\frac{1}{2}} \exp\left(-\frac{\Delta}{2\hbar^2}p^2\right),
\end{equation}
with some scale $\Delta$.
The separability condition implies \( \Delta x \sim \Delta^{1/2} \). For simplicity, setting \( \Delta x = (\Delta/2)^{1/2} \) yields
\begin{equation}
\phi(p) = \left( \frac{2}{\pi} \right)^{\frac{1}{4}} e^{-k^2} \left( \frac{\Delta x}{\hbar} \right)^{\frac{1}{2}},
\end{equation}
where \( \Delta x \) is the position uncertainty.

For initial states with even momentum-space wave functions and zero average momentum, we can express the position-space wave function as
\begin{equation}
\psi_{\Delta x}(x) = \frac{(\Delta x)^{\frac{1}{2}}}{\hbar} \int_{-\infty}^{\infty} \frac{\d p}{\sqrt{2\pi}} f\left( \frac{|p| \Delta x}{\hbar} \right) e^{ipx / \hbar},
\end{equation}
and the path length expectation
\begin{align}
\langle \Delta l \rangle = \int \d x ~ |x| \left| \psi_{\Delta x}(x, t) \right|^{2} 
= \Delta x \int \d y ~ |y| \left| F\left( y, b \right) \right|^{2},
\end{align}
where \( b \equiv \hbar t / [2m (\Delta x)^2] \) and
\begin{equation}
F(y, b) = \int_{-\infty}^{\infty} \frac{\d k}{\sqrt{2\pi}} f(|k|) e^{i k y - i k^{2} b}.
\end{equation}

Assuming a constant \( b \), we obtain the scaling relation:
\begin{equation}
\langle \Delta l \rangle \propto \Delta x \propto \frac{\hbar t}{m \Delta x},
\end{equation}
which yields Hausdorff dimension $d=2$. For states with average momentum $p_\text{av}$:
\begin{equation}
\psi_{\Delta x}(x) = \frac{(\Delta x)^{\frac{1}{2}}}{\hbar} \int_{-\infty}^{\infty} \frac{\d p}{\sqrt{2\pi}} f\left( \frac{|p| \Delta x}{\hbar} \right) e^{i(p + p_\text{av})x / \hbar},
\end{equation}
the path length becomes:
\begin{equation}
\langle \Delta l \rangle = \frac{|p_\text{av}| t}{m} \int \d y \left| \frac{p_\text{av}}{|p_\text{av}|} + \frac{\hbar y}{2|p_\text{av}| \Delta x \, b} \right| |F(y, b)|^{2}.
\end{equation}

Scale-independent Hausdorff length requires $d=1$ when resolving distances much larger than the de Broglie wavelength ($\Delta x \gg \hbar / |p_{\text{av}}|$), and $d=2$ when resolving much smaller distances ($\Delta x \ll \hbar / |p_{\text{av}}|$). Between these classical and quantum limits, $d$ remains ill-defined due to its strong $\Delta x$-dependence.

\section{Measurement Effects on Quantum paths and Hausdorff Dimension}

As established in previous discussions, periodic detections performed at intervals of $t$ do not constitute genuine quantum measurements. A complete quantum measurement requires several components: the target quantum system; a physical measurement apparatus; dynamical coupling to define the measurement basis; acquisition of the measurement outcomes; and, crucially, a state projection that induces backreaction and transitions of the system to a new quantum state.

We model sequential instantaneous position measurements at discrete times $t_r = r\tau$ ($r = 1, \dots, n$) using measurement paradigm developed in \cite{Caves:1987}. The quantum system couples sequentially to measurement apparatuses (termed ``meters''), where the $r^{\mathrm{th}}$ meter has canonical conjugate variables $\hat{\bar{x}}_r$ and $\hat{\bar{p}}_r$. Each meter serves as the initial stage of a macroscopic measurement chain. Assuming negligible meter free Hamiltonians, the composite system evolves under:

\begin{equation}
\hat{H} = \hat{H}_0 + \sum_{r=1}^{n} \delta(t - r\tau) \hat{x} \hat{\bar{p}}_r,
\label{eq:Hamiltonian}
\end{equation}
where $\hat{H}_0$ is the system's free Hamiltonian.

The time-dependent Hamiltonian \eqref{eq:Hamiltonian} violates closed-system dynamics, reflecting measurement-induced perturbations. The $\delta$-function coupling represents idealized instantaneous interactions. While generalized measurement theories extend beyond this framework, \eqref{eq:Hamiltonian} provides a basic, foundational model.

Assuming measurement homogeneity, we analyze a representative $r^{\mathrm{th}}$ measurement. The corresponding meter initializes in a Gaussian pure state
\begin{equation}
\langle \bar{x}_{r} | \Upsilon_{r} \rangle=\Upsilon(\bar{x}_{r})=(\pi\sigma)^{-\frac{1}{4}}\exp(-\bar{x}_{r}^{2}/2\sigma).
\end{equation}

Let $\hat{\rho}_{r}(t_{r}-)$ denote the pre-measurement system state at $t_r = r\tau$, implicitly conditioned on prior outcomes. Following the $\delta$-interaction in \eqref{eq:Hamiltonian}, the joint system-meter state is:
\begin{equation}
\hat{R}^{(r)} = e^{-i\hat{x}\hat{\bar{p}}_{r}/\hbar} \left( \ket{\Upsilon_{r}}\bra{\Upsilon_{r}} \otimes \hat{\rho}_{r}(t_{r}-) \right) e^{i\hat{x}\hat{\bar{p}}_{r}/\hbar}.
\label{eq:joint_state}
\end{equation}

The system measurement operator
\begin{equation}
\hat{\Upsilon}(\bar{x}_{r}) \equiv \bra{\bar{x}_{r}} e^{-i\hat{x}\hat{\bar{p}}_{r}/\hbar} \ket{\Upsilon_{r}}
\label{eq:measurement_op}
\end{equation}
equivalently expresses the position-translated wave function:
\begin{equation}
\hat{\Upsilon}(\bar{x}_{r}) = (\pi\sigma)^{-\frac{1}{4}} \exp\left[ -(\bar{x}_{r} - \hat{x})^{2}/(2\sigma) \right].
\label{eq:translated_wavefn}
\end{equation}

\subsection{Nonselective evolution and continuous position measurement}

Our objective is to derive a master equation for the system density operator $\hat{\rho}(t)$ when measurement outcomes are not recorded. We define its time derivative as
\begin{align}
\frac{\text{d}\hat{\rho}(t)}{\text{d}t} &:=\lim_{\tau \rightarrow 0} \frac{\hat{\rho}(t_n+) - \hat{\rho}(t_{n-1}+)}{\tau},
\label{eq:4.1}
\end{align}
where $\hat{\rho}(t_n+)$ and $\hat{\rho}(t_{n-1}+)$ are the \textit{nonselective} density operators immediately following the $n$\textsuperscript{th} and $(n-1)$\textsuperscript{th} measurements, respectively. For $n$ sequential measurements with interval $\tau$, the free evolution between measurements is governed by $
\hat{U}(\tau) = e^{-i \hat{H}_0 \tau / \hbar}$. The time derivative can be expanded into
\begin{align}
\frac{\text{d}\hat{\rho}(t)}{\text{d}t} &= \lim_{\tau \rightarrow 0} \Bigg[ \frac{1}{\tau} \int_{-\infty}^{\infty} d\bar{x} \, \hat{\Upsilon}(\bar{x}) \hat{U}(\tau) \hat{\rho}(t_{n-1}+) \hat{U}^{\dagger}(\tau) \hat{\Upsilon}^{\dagger}(\bar{x}) \nonumber \\
&\qquad - \frac{1}{\tau} \hat{\rho}(t_{n-1}+) \Bigg].
\label{eq:4.2}
\end{align}
We assume that the initial density operator of the system takes the form $\hat{\rho}_r(0) = \ket{\psi_r(0)}\bra{\psi_r(0)}$, where $\ket{\psi_r(0)}$ is a pure state characterized by a Gaussian wave function
\begin{equation}
\begin{split}
\psi(x,0) = {} & \left[\pi\Delta(0)\right]^{-\frac{1}{4}} \\
& \times \exp\left( -\frac{1 - i\epsilon(0)}{2\Delta(0)} [x - a(0)]^2 + \frac{i}{\hbar} b(0) x \right).
\label{psi}
\end{split}
\end{equation}
Taking the continuous limit ($ \tau \rightarrow 0 $, $ \sigma \rightarrow \infty $, with $ D = \sigma\tau = \mathrm{const.} $), we obtain the master equation \cite{Caves:1987}
\begin{equation}
\frac{\text{d}\hat{\rho}(t)}{\text{d}t} = -\frac{i}{\hbar} [\hat{H}_{0}, \hat{\rho}(t)] - \frac{1}{4D} [\hat{x}, [\hat{x}. \hat{\rho}(t)]].
\label{main}
\end{equation}

Here we need to clarify the definition of the time interval. When we are discussing the discrete case, the time interval should be $\tau$, during which the system undergoes free evolution for time $\tau$ followed by an instantaneous measurement process. However, when we take the continuous limit, $\tau \rightarrow 0$, the physical quantity that plays the role of time becomes $t$. Only when $D \rightarrow \infty$ do $\tau$ and $t$ become identical, which can also be seen from the time evolution of the wave packet width discussed later.

The master equation demonstrates that the evolution of the system's density matrix is comprised of two contributions: the first term on the right-hand side represents the \textit{intrinsic dynamics}, while the second term corresponds to the decay of off-diagonal elements—the \textit{decoherence term}. Fortunately, since we compute the expectation value of the absolute position \( |x| \), the contribution of decoherence is not explicitly present in the expression for $\langle |x| \rangle$, but is instead contained in the time evolution of the wave packet width. Without loss of generality, we express it as
\begin{align}
\langle |x| \rangle 
&= \int_{-\infty}^{\infty} \d x \, \frac{|x|}{\sqrt{\pi \Delta(t)}} 
\exp\left[ -\frac{1}{\Delta(t)} \left( x - a(t) \right)^{2} \right] \nonumber \\
&= |a(t)| \cdot \mathrm{erf}\left( \frac{|a(t)|}{\sqrt{\Delta(t)}} \right) 
+ \sqrt{\frac{\Delta(t)}{\pi}} \exp\left( -\frac{a(t)^{2}}{\Delta(t)} \right),
\label{eq:abs_expectation}
\end{align}
where $\mathrm{erf}(z)$ is the error function, defined as
\begin{equation*}
\mathrm{erf}(z) := \frac{2}{\sqrt{\pi}} \int_{0}^{z} \d t ~e^{-t^{2}}. 
\end{equation*}

From the master equation, for any operator $\hat{O}$, its expectation value $\langle \hat{O} \rangle \equiv \operatorname{Tr}(\rho(t) \hat{O})$ satisfies
\begin{equation}
\frac{\d\langle \hat{O} \rangle}{\d t} = -\frac{i}{\hbar} \langle [\hat{O}, \hat{H}_0] \rangle - \frac{1}{4D} \langle [\hat{x}, [\hat{x}, \hat{O}]] \rangle.
\label{eq:master}
\end{equation}

Thus we have
\begin{align}
\frac{\d\langle x \rangle}{\d t} &= \frac{\langle p \rangle}{m}, \\
\frac{\d\langle p \rangle}{\d t} &= 0.
\label{eq:derivatives}
\end{align}

Therefore, the evolution of the parameter $a(t)$ is given by the linear equation
\begin{equation}
a(t) = a(0) + b(0)t,
\label{eq:parameter_evolution}
\end{equation}
where $b(0)$ corresponds to the average momentum $p_{\text{av}}$ previously discussed. This also implies that the measurement process itself does not alter the motion of the wave packet center. When $b(0)$ is sufficiently large, we maintain the scaling relationships
\begin{align}
\langle |x| \rangle \sim t \quad \text{and} \quad d = 1.
\end{align}

When initial conditions satisfy $a(0) = b(0) = 0$, the expectation value becomes
\begin{align}
\langle |x| \rangle = \sqrt{\frac{\Delta(t)}{\pi}},
\end{align}
in which case we must compute the wave packet dispersion $\Delta(t)$. From Eq. \eqref{eq:master}, we have
\begin{align}
\frac{\d\langle x^2 \rangle}{\d t} &= \frac{1}{m} \langle xp + px \rangle, \\
\frac{\d\langle p^2 \rangle}{\d t} &= \frac{\hbar^2}{2D}, \\
\frac{\d\langle xp + px \rangle}{\d t} &= \frac{2}{m} \langle p^2 \rangle.
\end{align}
With the definitions of \( \sigma_x^2 \), \( \sigma_p^2 \) and \( \sigma_{xp} \) as follows,
\begin{align}
\sigma_x^2 &:= \langle (x - \langle x \rangle)^2 \rangle\nonumber, \\ \nonumber
\sigma_p^2 &:= \langle (p - \langle p \rangle)^2 \rangle \nonumber, \\
\sigma_{xp} &:= \frac{1}{2} \langle xp + px \rangle - \langle x \rangle \langle p \rangle \nonumber,
\end{align}
we obtain their time derivatives
\begin{align}
\frac{\d \sigma_x^{2}}{\d t} &= \frac{2}{m} \sigma_{xp}, \\
\frac{\d \sigma_p^{2}}{\d t} &= \frac{\hbar^2}{2D}, \\
\frac{\d \sigma_{xp}}{\d t} &= \frac{1}{m} \sigma_{p}^{2}.
\end{align}
Combining the three equations above, we have
\begin{equation}
\sigma_{x}^{2}(t) = \sigma_{x}^{2}(0) + \frac{2}{m} \sigma_{xp}(0) t + \frac{\sigma_{p}^{2}(0)}{m^{2}} t^{2} + \frac{\hbar^{2}}{6m^{2}D} t^{3}
\label{eq:sigma_x_t}
\end{equation}
From the initial state wave function \eqref{psi}, we can obtain
\begin{align}
\sigma_{x}^{2}(0) &= \frac{\Delta(0)}{2}, \label{eq:initial_sigma_x} \\
\sigma_{p}^{2}(0) &= \frac{\hbar^{2}(1 + \epsilon(0)^{2})}{2\Delta(0)}, \label{eq:initial_sigma_p} \\
\sigma_{xp}(0) &= \frac{\hbar \epsilon_{0}}{2}. \label{eq:initial_sigma_xp}
\end{align}
Then we have
\begin{equation}
\Delta(t) = \Delta(0) + \frac{2\hbar \epsilon(0)}{m} t + \frac{\hbar^{2}(1 + \epsilon(0)^{2})}{m^{2}\Delta(0)} t^{2} + \frac{\hbar^{2}}{3m^{2}D} t^{3}.
\label{eq:Delta_t}
\end{equation}
As previously stated, setting \( \Delta x = \left( \frac{\Delta(0)}{2} \right)^{\frac{1}{2}} \), we obtain
\begin{equation}
\begin{split}
\langle |x| \rangle = \frac{\Delta x}{\sqrt{\pi}} \Biggl( & 2 + \frac{2\hbar\epsilon(0) t}{m(\Delta x)^2} + \frac{\hbar^2(1+\epsilon(0)^2) t^2}{2m^2(\Delta x)^4} \\
& + \frac{\hbar^2 t^3}{3D m^2(\Delta x)^2} \Biggr)^{\frac{1}{2}}
\label{eq:position_expectation}.
\end{split}
\end{equation}

When \( D \rightarrow \infty \), if we take the constant \( b \equiv \hbar t / [2m (\Delta x)^2] \), the expression reduces to the result in \cite{Caves:1987} and we have $d=2$. Conversely, when \( D \rightarrow 0 \), the dominant term becoming \( (\Delta x)^3 \), and \( d \rightarrow 0 \) in this limit.

\subsection{Selective evolution and feedback control}
The previous analysis focuses on the nonselective evolution, which describes the evolution of the system's reduced density matrix and corresponding operator expectation values during system-meter interactions. In contrast, \textit{selective evolution} incorporates wave function collapse after each measurement, thus corresponding to a quantum jump from one pure state to another. Given an initial state $\hat{\rho}(0)$ and an outcome sequence $\{\bar{x}_r\}_{r=1}^n$, the state after $n$ measurements ($t_n = n\tau$) is
\begin{equation}
\begin{split}
\hat{\rho}(\{\bar{x}_r\}, t_n+) & = \frac{1}{P(\{\bar{x}_r\})} \left[ \prod_{r=1}^n \hat{\Upsilon}(\bar{x}_r) \hat{U}(\tau) \right] \\
& \quad \times \hat{\rho}(0) \left[ \prod_{r=1}^n \hat{\Upsilon}(\bar{x}_r) \hat{U}(\tau) \right]^\dagger,
\end{split}
\label{eq:final_state}
\end{equation}
with joint outcome probability
\begin{equation}
P(\{\bar{x}_r\}) = \tr \left\{ \left[ \prod_{r=1}^n \hat{\Upsilon}(\bar{x}_r) \hat{U}(\tau) \right]^\dagger \left[ \prod_{r=1}^n \hat{\Upsilon}(\bar{x}_r) \hat{U}(\tau) \right] \hat{\rho}(0) \right\}.
\label{eq:joint_probability}
\end{equation}

Consistent with the quantum trajectory framework, we assume $\ket{\psi_r(t)}$ is a pure state similar to \eqref{psi}
\begin{equation}
\begin{split}
\psi_r(x,t) = {} & e^{i\phi_r(t)} \left[\pi\Delta_r(t)\right]^{-\frac{1}{4}} \\
& \times \exp\left( -\frac{1 - i\epsilon_r(t)}{2\Delta_r(t)} [x - a_r(t)]^2 + \frac{i}{\hbar} b_r(t) x \right),
\end{split}
\end{equation}
where the parameters $\{a_r, b_r, \Delta_r, \epsilon_r, \phi_r\}$ are smooth functions of time during unitary evolution within the interval $(t_{r-1}, t_r)$. These determine the expectation values of the relevant observables.

We shall focus on parameter values immediately after the $(r-1)$-th measurement and just before the $r$-th measurement. We will adopt some concise notations:
\begin{itemize}
\item Unadorned parameters denote post-$(r-1)$-measurement values: $\Delta_r \equiv \Delta_r(t_{r-1}+)$;
\item Primed parameters denote pre-$r$-measurement values: $\Delta_r^\prime \equiv \Delta_r(t_r-)$.
\end{itemize}

Unitary evolution relates the primed parameters to the unprimed parameters via
\begin{align}
a_r^\prime &= a_r + b_r \tau / m, \label{eq:3.18a} \\
b_r^\prime &= b_r, \label{eq:3.18b} \\
\Delta_r^\prime &= \Delta_r + \frac{1 + \epsilon_r^2}{\Delta_r} \left( \frac{\hbar \tau}{m} \right)^2 + 2\epsilon_r \frac{\hbar \tau}{m}, \label{eq:3.19a} \\
\epsilon_r^\prime &= \epsilon_r + \frac{1 + \epsilon_r^2}{\Delta_r} \frac{\hbar \tau}{m}. \label{eq:3.19b}
\end{align}

Following the $r$-th measurement, the system collapses to a pure state
\begin{equation}
\ket{\psi_{r+1}(t_r+)} = \frac{\hat{\Upsilon}(\bar{x}_r) \ket{\psi_r(t_r-)}}{\left[ P(\bar{x}_r) \right]^{\frac{1}{2}}}, \label{eq:3.22}
\end{equation}
whose Gaussian wave function $\psi_{r+1}(x, t_r+) \propto \Upsilon(\bar{x}_r - x) \psi_r(x, t_r-)$ is characterized (up to a phase) by updated parameters
\begin{align}
a_{r+1} &= a_r^\prime + \frac{C_r - 1}{C_r} (\bar{x}_r - a_r^\prime), \label{eq:3.23a} \\
b_{r+1} &= b_r^\prime + \hbar \frac{\epsilon_r^\prime}{\Delta_r^\prime} \frac{C_r - 1}{C_r} (\bar{x}_r - a_r^\prime), \label{eq:3.23b} \\
\Delta_{r+1} &= \Delta_r^\prime / C_r, \label{eq:3.24a} \\
\epsilon_{r+1} &= \epsilon_r^\prime / C_r, \label{eq:3.24b}
\end{align}
where the contraction factor
\begin{equation}
C_r \equiv 1 + \Delta_r^\prime / \sigma \geqslant 1 \label{eq:3.25}
\end{equation}
quantifies the reduction in position variance due to measurement. If we assume the same value of $\Delta$ and $\epsilon$ after each measurement, then the mean position and mean momentum change according to
\begin{align}
a_{r+1} - a_{r} &= b_{r}\frac{\tau}{m} + \frac{C-1}{C}(\bar{x}_{r}-a_{r}^{\prime}) \\
b_{r+1} - b_{r} &= \frac{m}{\tau}\frac{(C-1)^{2}}{C(C+1)}(\bar{x}_{r}-a_{r}^{\prime})= \frac{\hbar}{C^{\frac{1}{2}}\sigma}(\bar{x}_{r}-a_{r}^{\prime}),
\end{align}
where $C \equiv 1 + \Delta^{\prime} / \sigma \geq 1$.

However, we now confront a fundamental challenge: the measurement outcome $\bar{x}_r$ is not a deterministic value but corresponds to a probability distribution. For each measurement instance, $\bar{x}_r$ can assume entirely different values governed by quantum randomness. 

Since each measurement induces unpredictable ``jumps'' in the mean position and momentum values due to measurement outcome stochasticity, after multiple measurements, these parameters typically undergo significant deviation from their initial values. In any physically realizable measurement scenario, such behavior is precluded; the measured system must remain confined within the laboratory with near-unity probability. To explicitly embed this physical constraint into our measurement formalism, we introduce feedback forces that counteract the mean position and momentum jumps. Formally, this necessitates modifying the operational specification of the measuring apparatus.

The key idea is to introduce a feedback force after each measurement to counteract the jumps in position and momentum coordinates, thereby canceling the uncertainty in $\bar{x}_r$. This can be implemented via a displacement operator
\begin{align}
\hat{D}(\overline{\boldsymbol{x}}_{r}) := \exp\left[ -\frac{i}{\hbar} \tau \overline{\boldsymbol{x}}_{r} (\gamma_{1}\hat{\boldsymbol{x}} - \gamma_{2}\hat{\boldsymbol{p}}) \right],
\end{align}
where
\begin{align}
\gamma_{1} \equiv \hbar / D \equiv m / (2 t_{c}^{2}), \quad
\gamma_{2} \equiv 1 / t_{c}.
\end{align}

Thus we have the discrete difference equations
\begin{align}
a_{r+1} - a_{r} &= \left( \frac{b_{r}}{m} - \frac{a_{r}}{t_{c}} \right) \tau \label{eq:diff_a}, \\
b_{r+1} - b_{r} &= -\frac{m }{2t_c^2}a_{r} \tau.
\label{eq:diff_b}
\end{align}

In the continuous limit $\tau \to 0$, the finite differences converge to time derivatives
\begin{equation*}
\frac{a_{r+1} - a_{r}}{\tau} \xrightarrow{\tau \to 0} \frac{da}{dt}, \quad
\frac{b_{r+1} - b_{r}}{\tau} \xrightarrow{\tau \to 0} \frac{db}{dt},
\end{equation*}
yielding the coupled differential equations
\begin{align}
\frac{\d a}{\d t} &= \frac{b}{m} - \frac{a}{t_{c}} \label{eq:da_dt}, \\
\frac{\d b}{\d t} &= -\frac{m }{2t_c^2}a 
\label{eq:db_dt}
\end{align}

Differentiating the above two equations with respect to time, we obtain the second-order harmonic oscillator equation
\begin{align}
\frac{\d{}^{2} a}{\d t^{2}} + \frac{1}{t_{c}} \frac{\d a}{\d t} + \frac{1}{2t_c^2} a &= 0 \label{eq:osc_a}, \\
\frac{\d{}^{2} b}{\d t^{2}} + \frac{1}{t_{c}} \frac{\d b}{\d t} + \frac{1}{2t_c^2} b &= 0 \label{eq:osc_b}.
\end{align}

These equations correspond precisely to a damped harmonic oscillator. For sufficiently large $t$, irrespective of initial conditions, the system evolves to a stationary state characterized by
\begin{align}
\langle x \rangle = 0, \quad
\langle p \rangle = 0,
\end{align}
while the expectation values of other operators approach constant values. Consequently, $\langle |x| \rangle$ becomes proportional to the position uncertainty $\Delta x$ and we always have $d=2$.

\section{Discussion}

In this work we investigate how physical quantum measurements disrupt the fractal geometry of quantum particle paths. Idealized models predict that quantum paths exhibit a universal Hausdorff dimension \(d = 2\) due to Heisenberg uncertainty principle, which causes the path length to diverge with finer spatial resolution. {However, although it was assumed that measurements occur at intervals of time $t$, calculations merely involved evaluating the expectation value of operators corresponding to the wave function evolution within a single time interval $t$. Genuine physical measurements did not actually take place.

In our work, we employ wave packets as an example to explicitly model the interaction between the system and measurement apparatus. We systematically analyze two distinct processes:
\begin{itemize}
    \item {Decoherence} -- where the system transitions from a pure state to a mixed state;
    \item {Quantum state collapse} -- where the system evolves from one pure state to another.
\end{itemize}
}
In nonselective evolution, measurements perturb the path roughness without wave function collapse. The master equation for the system's density matrix includes a decoherence term proportional to \(1/D\), where \(D\) is a parameter characterizing measurement strength (with \(D = \sigma\tau\) for meter uncertainty \(\sigma\) and time interval \(\tau\)). As \(D \to \infty\) (weak measurement), the Hausdorff dimension approaches the ideal \(d = 2\). However, as \(D \to 0\) (strong measurement), decoherence dominates, suppressing quantum fluctuations and reducing the dimension toward \(d = 0\). This shift arises because measurement backaction smooths out the path, diminishing its fractal character. For states with an average momentum \(p_{\text{av}}\), the dimension transitions from the classical value of \(d = 1\) at coarse resolutions to the quantum value \(d = 2\) at fine resolutions, but measurement effects blur this transition, making \(d\) strongly dependent on \(\Delta x\).

In selective evolution, measurements record outcomes, causing stochastic wave function collapses that lead to unstable trajectories with unpredictable jumps in position and momentum. To stabilize the paths, we introduce feedback control via a displacement operator that counteracts measurement-induced jumps. This feedback forces the system into a damped harmonic oscillator-like evolution, driving mean position and momentum to zero and restoring the Hausdorff dimension to \(d = 2\) regardless of initial conditions. Thus, feedback is essential for maintaining fractal properties in experimental settings.

{
This approach leads to a key conclusion: physical measurements significantly perturb the geometric structure of quantum paths. Consequently, the conventional understanding -- that the Hausdorff dimension transitions from $d=2$ to $d=1$ as momentum increases -- cannot be taken for granted. Instead, the contributions of decoherence and wave packet collapse during actual measurement processes must be rigorously accounted for. This work bridges quantum fractality theory with experimental measurement physics, highlighting how detectors reshape spacetime statistics at quantum scales.}

Future research should focus on extending this framework through several key generalizations. First, deeper connections between measurement theory and quantum gravity concepts—particularly minimal length scales and the generalized uncertainty principle \cite{Hossenfelder:2012jw}—warrant theoretical exploration (see related works in \cite{Nicolini:2010dj,Lake:2022hzr}). The dimensional alterations may also imply violations of Lorentz invariance \cite{AmelinoCamelia:2013zea,Collins:2004bp}, potentially enabling the detection of observable signatures through measurement \cite{Davies:2023zfq,Rinaldi:2007de,Rinaldi:2008qt,Husain:2015tna,Xu:2025smb}.

Second, while the current model studies non-relativistic quantum mechanics, transitioning from non-relativistic quantum mechanics to relativistic regimes also presents a crucial challenge; comparative analysis with established models like the Unruh-DeWitt detector would be particularly illuminating. Conventional Unruh-DeWitt detector frameworks exclusively address excitations and decoherence within two-level systems \cite{Davies,Nesterov:2020exl,Xu:2022juv,Xu:2023tdt}. By contrast, our methodology adopts Gaussian wavepacket to probe how quantum path decoherence modifies the Unruh effect.

Finally, since quantum gravity essentially operates in curved spacetimes, future research must incorporate measurements in that setting. For example, we can use detectors in environments such as AdS spacetime to rigorously test the holographic principle through the AdS/CFT correspondence \cite{Ng:2018drz,Ahmed:2020fai,Ahmed:2023uem,Pitelli:2021oil}. We can also consider the exchange of energy and information between detectors and the detection objects \cite{Xu:2021buk,Xu:2024xlx,Xu:2024ztq}. These directions constitute the key next step in unifying quantum measurement physics with gravitational phenomena.

\begin{acknowledgments}
Hao Xu thanks National Natural Science Foundation of China (No.12205250) for funding support.
\end{acknowledgments}



\end{document}